\documentclass[12pt]{iopart}

\usepackage{graphicx}
\usepackage{color}
\usepackage{amssymb}
\usepackage{mathrsfs}

\begin{document}

\title{Quantifying the non-ergodicity of scaled Brownian motion}

\author{Hadiseh Safdari$^\dagger$, Andrey G. Cherstvy$^\ddagger$, Aleksei V.
Chechkin$^{\ddagger,\P}$, Felix Thiel$^\flat$, Igor M. Sokolov$^\flat$,
and Ralf Metzler$^{\ddagger,\sharp}$\thanks{E-mail:rmetzler@uni-potsdam.de}}
\address{$\dagger$ Department of Physics, Shahid Beheshti University, G.C.,
Evin, 19839 Tehran, Iran\\
$\ddagger$ Institute for Physics \& Astronomy, University of Potsdam, 14476
Potsdam-Golm, Germany\\
$\P$ Institute for Theoretical Physics, Kharkov Institute of Physics and
Technology, 61108 Kharkov, Ukraine\\
$\flat$ Institute for Physics, Humboldt-Universit{\"a}t zu Berlin, 12489 Berlin,
Germany\\
$\sharp$ Department of Physics, Tampere University of Technology, 33101 Tampere,
Finland}

\begin{abstract}
We examine the non-ergodic properties of scaled Brownian motion, a
non-stationary stochastic process with a time dependent diffusivity of
the form $D(t)\simeq t^{\alpha-1}$. We compute the ergodicity breaking
parameter EB in the entire range of scaling exponents $\alpha$, both
analytically and via extensive computer simulations of the stochastic Langevin
equation. We demonstrate that in the limit of long trajectory lengths $T$
and short lag times $\Delta$ the EB parameter as function of the scaling exponent
$\alpha$ has no divergence at $\alpha=1/2$ and present the asymptotes for EB in
different limits. We generalise the analytical and simulations results
for the time averaged and ergodic properties of scaled Brownian motion
in the presence of ageing, that is, when the observation of the system starts
only a finite time span after its initiation. The approach developed here
for the calculation of the higher time averaged moments of the
particle displacement can be applied to derive the ergodic properties
of other stochastic processes such as fractional Brownian motion.
\end{abstract}

\pacs{05.40.−a,02.50.-r,87.10.Mn}

\date{\today}

\section{Introduction} 
       
The non-Brownian scaling of the mean squared displacement (MSD) of a diffusing
particle of the power-law form \cite{bouchaud,metz00,franosch,metz14}
\begin{equation}
\label{msd}
\langle x^2(t)\rangle=2K_{\alpha}t^{\alpha}
\end{equation}
is a hallmark of a wide range of anomalous diffusion processes \cite{metz00,
metz14}. Equation (\ref{msd}) features the anomalous diffusion coefficient
$K_{\alpha}$ of physical dimension $\mathrm{cm}^2/\mathrm{sec}^{\alpha}$ and
the anomalous diffusion exponent $\alpha$. Depending on its magnitude we distinguish
subdiffusion ($0<\alpha<1$) and superdiffusion ($\alpha>1$). Interest in anomalous
diffusion processes was rekindled with the advance of modern spectroscopic methods,
in particular, advanced single particle tracking methods \cite{brauchle}. Thus,
subdiffusion was observed for the motion of biopolymers and submicron tracer
particles in living biological cells \cite{lene}, in complex fluids \cite{weiss},
as well as in extensive computer simulations of membranes \cite{membrane} or
structured systems \cite{godec}, among others \cite{franosch,metz14,soko15}.
Superdiffusion of tracer particles was observed in living cells due to
active motion \cite{elbaum}.

Anomalous diffusion processes characterised by the MSD (\ref{msd}) may originate
from a variety of distinct physical mechanisms \cite{bouchaud,franosch,metz14,
soko15,metz12,soko12}. These include a power-law statistic of trapping times in
the continuous time random walks (CTRWs) as well as related random energy models
\cite{metz14,soko15,metz12,soko12,mont65,metz13} and CTRW variants with correlated
jumps \cite{corr} or superimposed environmental noise \cite{noisy}. Other
models include random processes driven by Gaussian yet power-law correlated noise
such as fractional Brownian motion (FBM) \cite{fbm} or the fractional Langevin
equation \cite{fle}. Closely related to these models is the subdiffusive motion
on fractals such as critical percolation clusters \cite{yasmine}. Finally, among
the popular anomalous diffusion models we mention heterogeneous diffusion processes
with given space dependencies of the diffusion coefficient \cite{cher13} as well as
processes with explicitly time dependence diffusion coefficients, in particular, the
scaled Brownian motion (SBM) with power-law form $D(t)\simeq t^{\alpha-1}$ analysed
in more detail herein \cite{muni02,fuli13,soko14,jeon14}. Also combinations of space
and time dependent diffusivities were investigated \cite{fuli13,cher15b}. Space
and/or time dependent diffusivities were used to model experimental results for
smaller tracer proteins in living cells \cite{lang11} and anomalous diffusion in
biological tissues \cite{nick08} including brain matter \cite{novi11,novi14}.
In particular, SBM was used to describe fluorescence recovery after photobleaching
in various settings \cite{saxt01} as well as anomalous diffusion in various
biophysical contexts \cite{post12}. In other branches of physics SBM was used to
model turbulent flows  observed by Richardson \cite{rich26} as early as 1952 by
Batchelor \cite{batc52}. Moreover, the diffusion of particles in granular gases
with relative speed dependent restitution coefficients follow SBM \cite{bodr15b}.
We note that in the limiting case $D(t)\sim 1/t$ the resulting process is ultraslow
with a logarithmic growth of the MSD \cite{bodr15a} known from processes such as
Sinai diffusion \cite{sinai}, single file motion in ageing environments \cite{sf},
or granular gas diffusion with constant restitution coefficient \cite{bodr15a}.

In the following we study the ergodic properties of SBM in the Boltzmann-Khinchin
sense \cite{bolkhin},
finding that even long time averages of physical observables such as the MSD do not
converge to the corresponding ensemble average \cite{metz14,metz12,soko12,metz08}.
In particular we compute the ergodicity breaking parameter EB---characterising the
trajectory-to-trajectory fluctuations of the time averaged MSD---in the entire range
of the scaling exponents $\alpha$, both analytically and from extensive computer
simulations. We generalise the results for the ergodic properties of SBM in the
presence of ageing, when we start to evaluate the time average the MSD a finite time
span after the initiation of the system.

The paper is organised as follows. In section \ref{sec-observables} we summarise
the observables computed and provide a brief overview of the basic
properties of SBM. In section \ref{sec-model-simul} we describe the theoretical
concepts and numerical scheme employed in the paper. We present the main results
for the EB parameter of non-ageing and ageing SBM in detail in sections
\ref{sec-non-aged} and \ref{sec-aged}. In section \ref{sec-disc} we summarise our
findings and discuss their possible applications and generalisations.

\section{Observables and fundamental properties of scaled Brownian motion}
\label{sec-observables}

We define SBM in terms of the stochastic process \cite{metz14,muni02,soko14,
cher15b,metz15} 
\begin{equation}
\frac{dx(t)}{dt}=\sqrt{2D(t)}\times\zeta(t),
\end{equation}
where $\zeta(t)$ is white Gaussian noise with zero mean and unit amplitude
$\langle\zeta(t_1)\zeta(t_2)\rangle=\delta(t_1-t_2)$. The time dependent
diffusion coefficient is taken as
\begin{equation}
D(t)=\alpha K_\alpha t^{\alpha-1},
\label{eq-d-versus-alfa}
\end{equation}
where we require the positivity of the scaling exponent, $\alpha>0$. SBM is
inherently out of thermal equilibrium in confining external potentials
\cite{jeon14}. Let us
briefly outline the basic properties of the SBM process. The ensemble averaged
MSD of SBM scales anomalously with time in the form of equation (\ref{msd}).

Here and below we use the standard definition of the time averaged MSD
\cite{metz14,metz12} 
\begin{equation}
\label{eq-TAMSD}
\overline{\delta^2(\Delta)}=\frac{1}{T-\Delta}\int\limits_0^{T-\Delta}\Big[x(
t+\Delta)-x(t)\Big]^2dt,
\end{equation}
where $\Delta$ is the lag time, or the width of the window slid along the time
series in taking the time average (\ref{eq-TAMSD}). Moreover, $T$ is the total
length of the time series. We denote ensemble averages by the angular brackets
while time averages are indicated by the overline. Often, an additional average
of the form
\begin{equation}
\label{eatamsd}
\left<\overline{\delta^2(\Delta)}\right>=\frac{1}{N}\sum_{i=1}^N\overline{\delta
^2_i(\Delta)}
\end{equation}
is performed over $N$ realisations of the process, to obtain smoother curves. From
a mathematical point of view, this trajectory average allows the calculation of
the time averaged MSD for processes, which are not self-averaging \cite{metz14,
metz08}\footnote{That is, a sufficiently long time average is sufficient to
represent the whole ensemble.}
Both quantities (\ref{eq-TAMSD}) and (\ref{eatamsd}) are important in the
analysis of single particle trajectories measured in advanced tracking experiments
\cite{metz12}. For SBM the mean time averaged MSD (\ref{eatamsd}) grows as
\cite{jeon14}
\begin{equation}
\left<\overline{\delta^2(\Delta)}\right>=\frac{2K_\alpha\left[T^{\alpha+1}-\Delta^{
\alpha+1}-(T-\Delta)^{\alpha+1}\right]}{(\alpha+1)(T-\Delta)}.
\label{eq-sbm-tamsd}
\end{equation} 
In the limit $\Delta/T\ll1$, the time averaged MSD scales linearly with the lag
time, 
\begin{equation}
\left<\overline{\delta^2(\Delta)}\right>\sim2K_\alpha\frac{\Delta}{T^{1-\alpha}}.
\end{equation}
SBM is thus a weakly non-ergodic process in Bouchaud's sense \cite{weak}: the
ensemble and time averaged MSDs are disparate even in the limit of long observation
times $T$, $\lim_{T\to\infty}\overline{\delta^2(\Delta)}\neq\langle x^2(t)\rangle$
and thus violate the Boltzmann-Khinchin ergodic hypothesis, while the entire phase
space is accessible to any single particle. Moreover, the magnitude
of the time averaged MSD becomes a function of the trace length $T$. Analogous
asymptotic forms for the mean time averaged MSD (\ref{eatamsd}) are found in
subdiffusive CTRW processes \cite{metz08,hensalu} and heterogeneous diffusion
processes \cite{cher13}, see also the extensive recent review \cite{metz14}.
Note that also much weaker forms of non-ergodic behaviour exist for L{\'e}vy
processes \cite{lw}.

Another distinct feature of weakly non-ergodic processes of the subdiffusive CTRW
\cite{metz08} and heterogeneous diffusion type \cite{cher13} is the fact that time
averaged observables remain random quantities even in the long time limit and thus
exhibit a distinct scatter of amplitudes between individual realisations for a
given lag time. This irreproducibility due to the scatter of individual traces
$\overline{\delta^2(\Delta)}$ around their mean is described by the
ergodicity breaking parameter \cite{metz14,metz08,deng09,rytov}
\begin{equation}
\mathrm{EB}(\Delta)=\frac{\left<\left(\overline{\delta^2(\Delta)}\right)^2\right>-
\left<\overline{\delta^2(\Delta)}\right>^2}{\left<\overline{\delta^2(\Delta)}\right>
^2}=\frac{\mathcal{N}(\Delta)}{\mathcal{D}(\Delta)}=\left<\xi^2(\Delta)\right>-1,
\label{eq-eb-via-xi}
\end{equation}
where $\xi(\Delta)=\overline{\delta^2(\Delta)}\Big/\left<\overline{\delta^2(
\Delta)}\right>$. Moreover, we introduced the abbreviations $\mathcal{N}(\Delta)$
and $\mathcal{D}(\Delta)$ for the nominator and denominator of EB, respectively.
This notation will be used below. For Brownian motion in the limit $\Delta/T\to0$
the EB parameter vanishes linearly with $\Delta/T$ in the form \cite{metz14,deng09}
\begin{equation}
\mathrm{EB}_{\mathrm{BM}}(\Delta)=\frac{4\Delta}{3T}. 
\label{eq-eb-bm}
\end{equation}
In contrast to subdiffusive CTRW and heterogeneous diffusion processes, the EB
parameter of SBM vanishes in the limit $\Delta/T\to0$ and in this sense the time
averaged observable becomes
reproducible \cite{soko14,jeon14,metz15}. We demonstrate the small amplitude
scatter of SBM in figure \ref{fig-tamsd-aged}, for a detailed discussion see below.
We note that the scatter of the time averaged MSD of SBM around the ergodic value
$\xi=1$ becomes progressively asymmetric for smaller $\alpha$ values and in later
parts of the time averaged trajectories, see Fig.~6 of reference \cite{metz14}. 
In the following we derive the exact analytical results for the EB parameter of
SBM and support these results with extensive computer simulations. Moreover we
extend the analytical and computational analysis of the EB parameter to the case
of the ageing SBM process when we start evaluating the time series $x(t)$
at the time $t_a>0$ after the original initiation of the system at $t=0$
\cite{metz15}.

\begin{figure} 
\includegraphics[width=8cm]{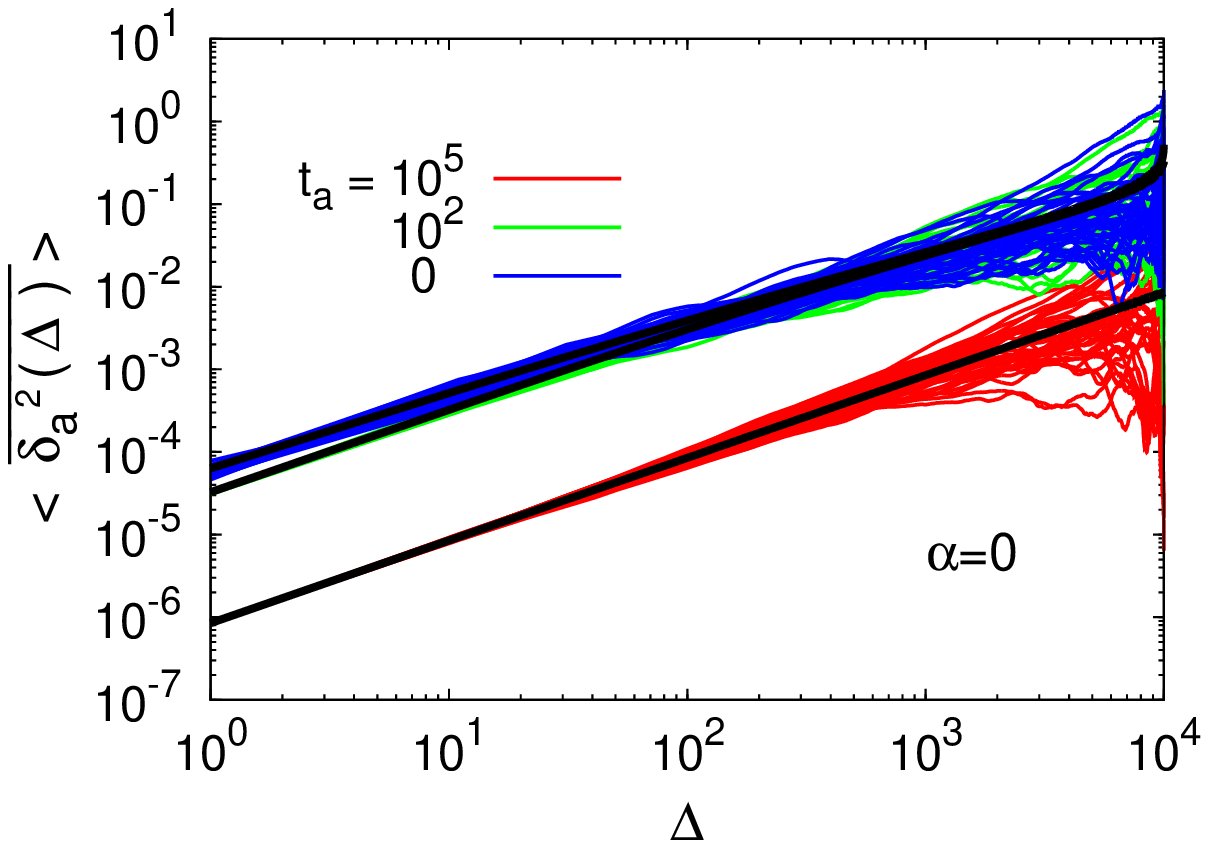}
\includegraphics[width=8cm]{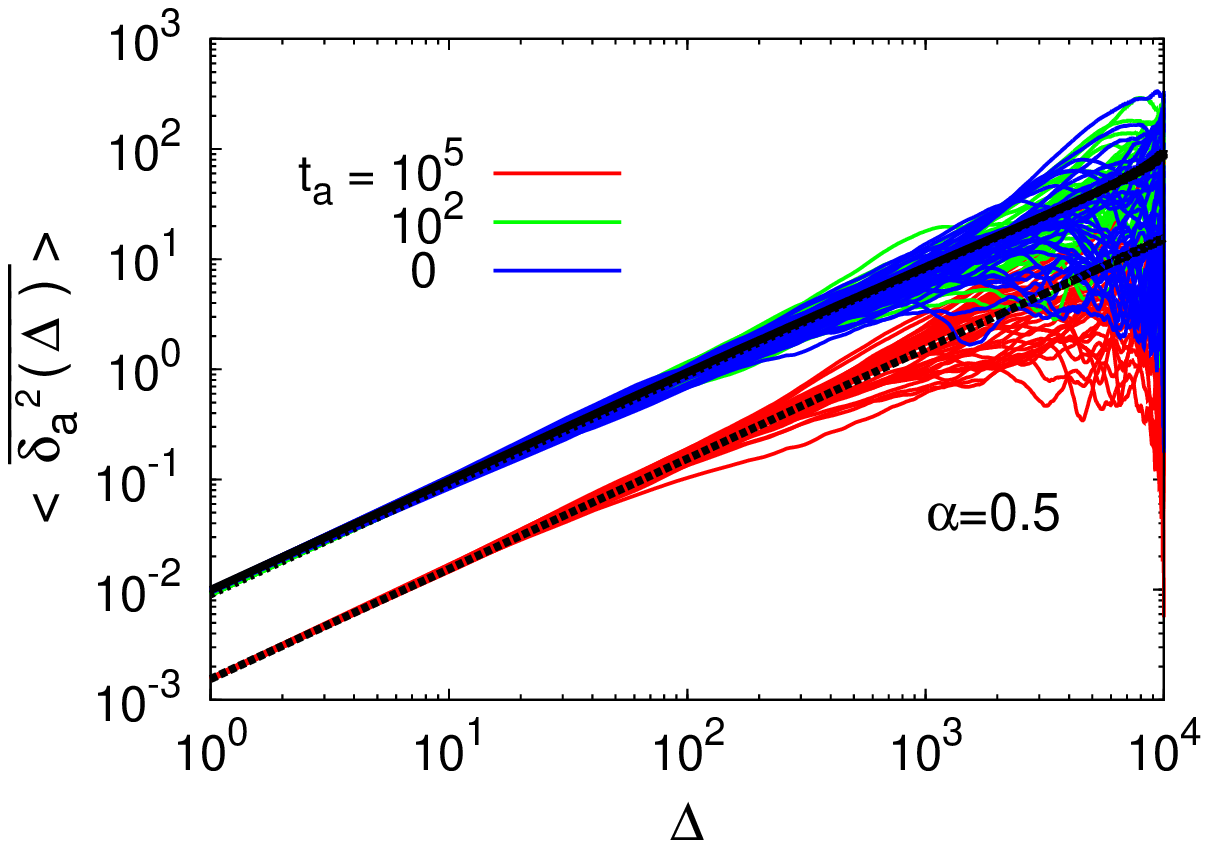}
\includegraphics[width=8cm]{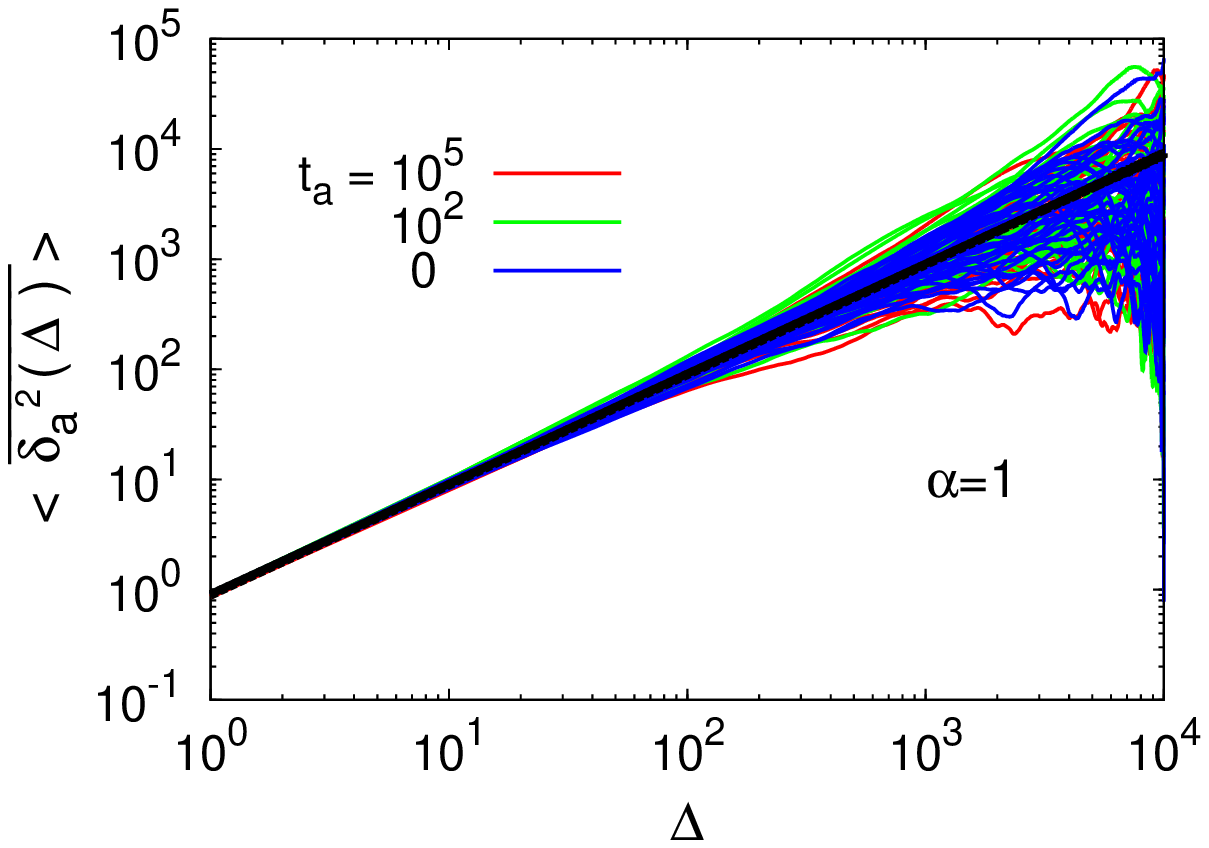}
\includegraphics[width=8cm]{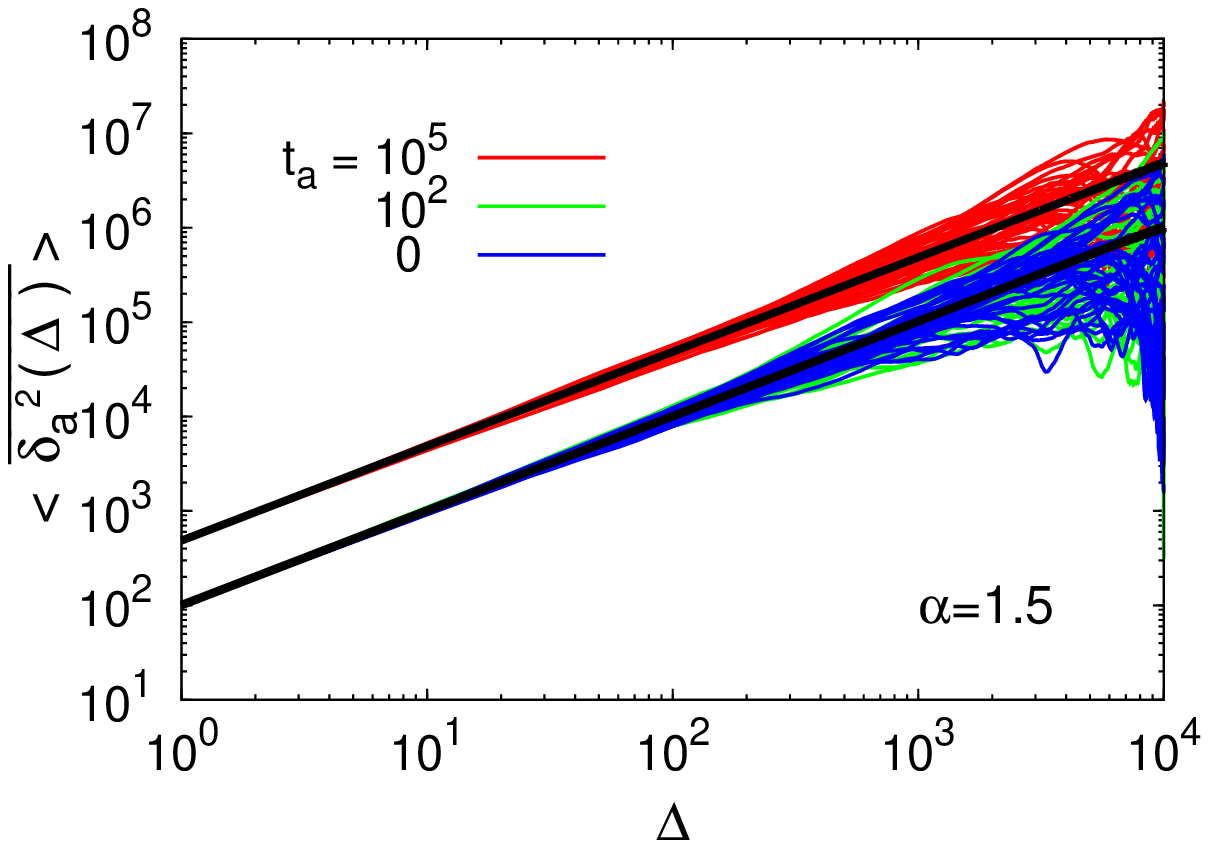}
\caption{Time averaged MSD of SBM as function of the lag time $\Delta$ for 
several values of the scaling exponents $\alpha$ and ageing times $t_a$. The
asymptotic behaviour of equation (\ref{eq-aged-sbm-delta-2-T-Delta}) is shown by
the black solid lines. Parameters: $T=10^4$, $t_a=0$, $10^2$, $10^5$, and
$N=100$ traces are shown.} 
\label{fig-tamsd-aged}
\end{figure}

The time averaged MSD of an ageing stochastic process is defined as \cite{metz13}
\begin{equation}
\label{eq-TAMSD-aged}
\overline{\delta^2_a(\Delta)}=\frac{1}{T-\Delta}\int_{t_a}^{t_a+T-\Delta}
\Big[x(t+\Delta)-x(t)\Big]^2dt
\end{equation}
and thus again involves the observation time $T$. The properties ageing SBM were
considered recently \cite{metz15}. The mean time averaged MSD becomes
\begin{eqnarray}
\nonumber
\left<\overline{\delta^2_a(\Delta)}\right>&=&\frac{2K_{\alpha}}{(\alpha+1)
(T-\Delta)}\Big[(T+t_a)^{\alpha+1}-(t_a+\Delta)^{\alpha+1}\\
&&-(T+t_a-\Delta)^{\alpha+1}+t_a^{\alpha+1}\Bigg].
\label{eq-aged-sbm-delta-2-T-Delta}
\end{eqnarray}
The ratio of the aged versus the non-ageing time averaged MSD in the limit $\Delta\ll
t_a,T$ has the asymptotic form \cite{metz15}
\begin{equation}
\Lambda_\alpha(t_a/T)=\frac{\left<\overline{\delta^2_a(\Delta)}\right>}{\left<
\overline{\delta^2(\Delta)}\right>}\sim(1+t_a/T)^\alpha-(t_a/T)^\alpha.
\end{equation}
This functional form is identical to that obtained for subdiffusive CTRWs
\cite{metz13} and heterogeneous diffusion processes \cite{andrey_age}.
The factor $\Lambda_\alpha(z)$ quantifies the respective depression and enhancement
of the time averaged MSD for the cases of ageing sub- and superdiffusive SBM. 

\begin{figure}
\begin{center}
\includegraphics[width=7.4cm]{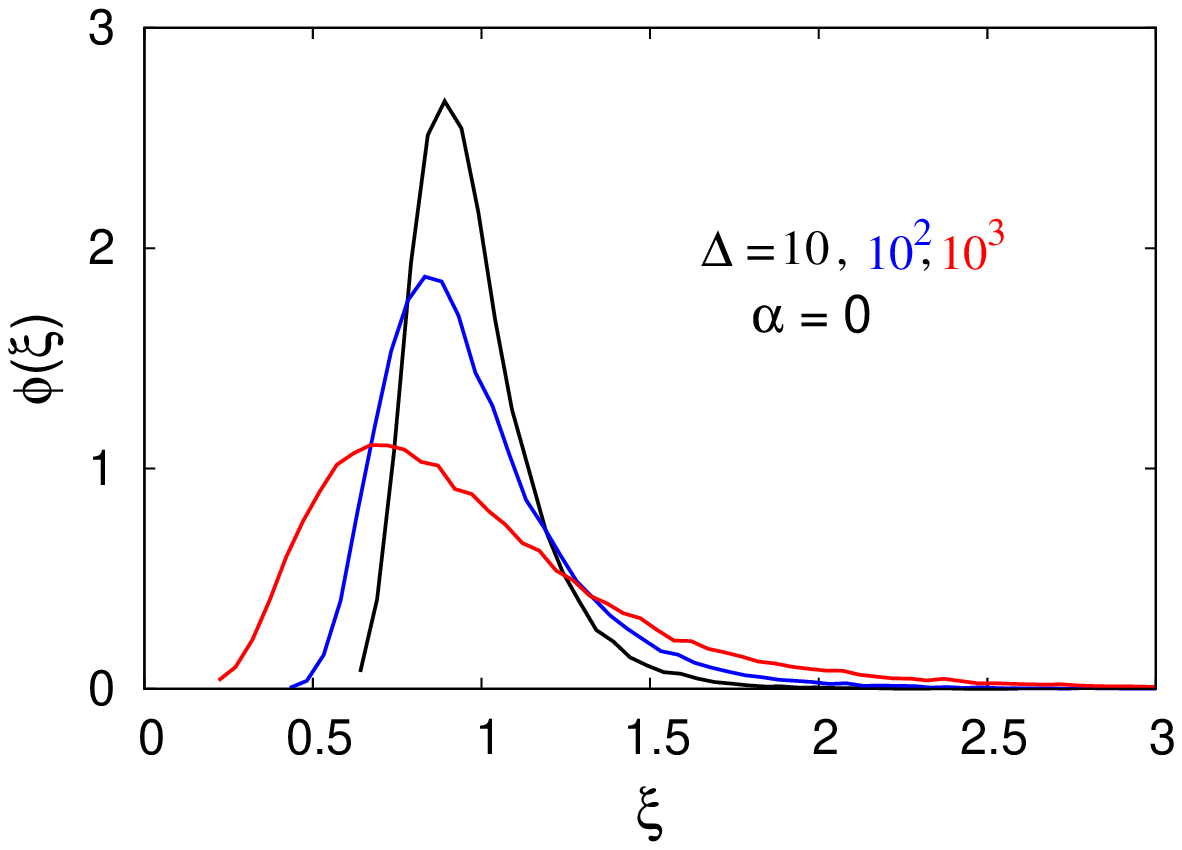}
\includegraphics[width=7.4cm]{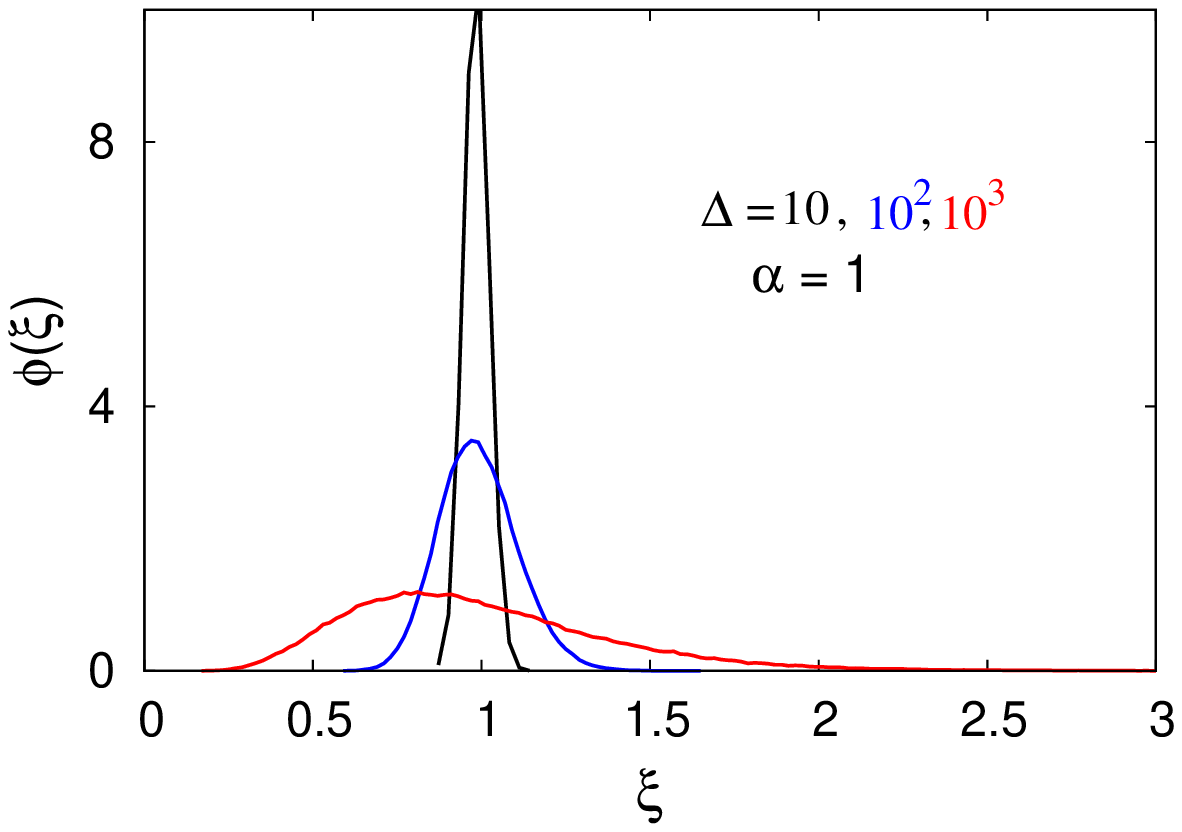}
\includegraphics[width=7.4cm]{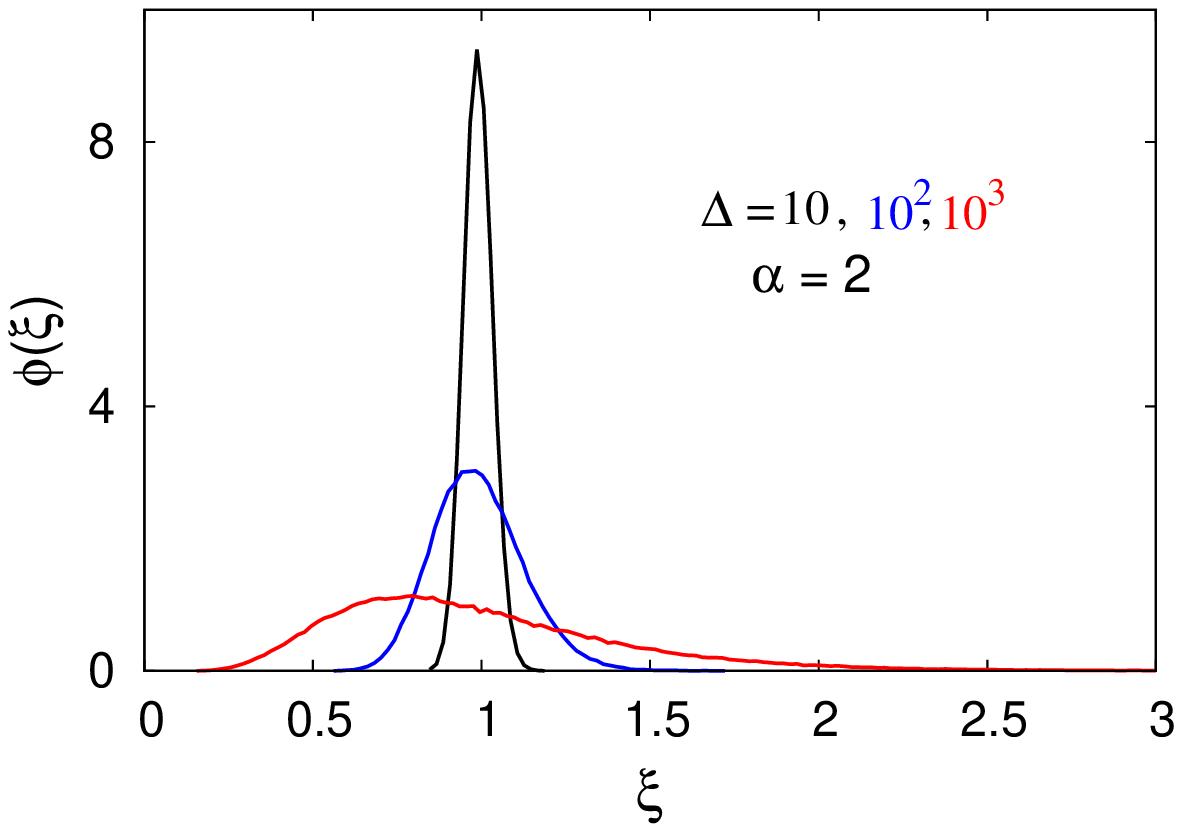}
\includegraphics[width=7.4cm]{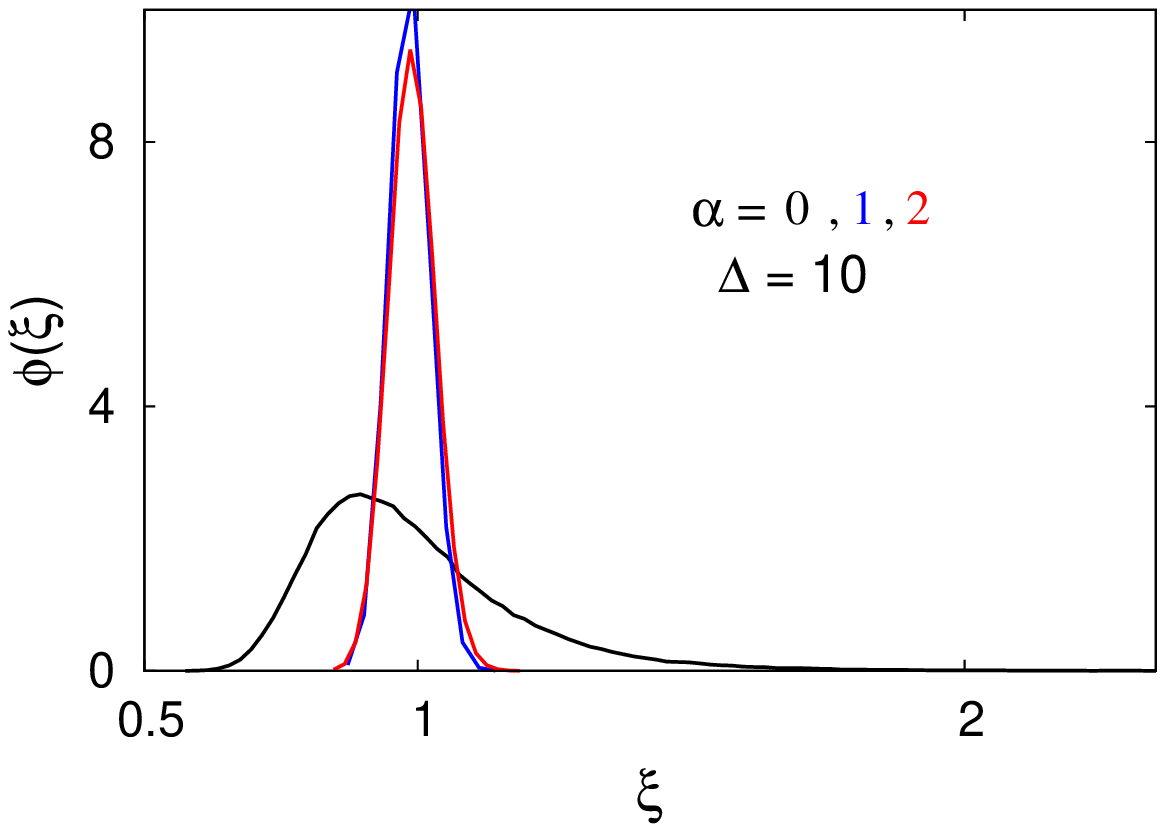}
\end{center}
\caption{Distribution $\phi(\xi)$ of the relative amplitude $\overline{\delta^2(
\Delta)}\Big/\left<\overline{\delta^2(\Delta)}\right>$ of the time averaged MSD
traces for SBM processes with different scaling exponents $\alpha$ as indicated
in the panels. As expected, the spread grows and the distribution becomes more
leptokurtic at longer lag times $\Delta$. For progressively larger values of the
scaling exponent $\alpha$ the spread of time averaged MSD decreases but stays
asymmetric with a longer tail at larger $\overline{\delta^2}$ values. In
particular, for $\alpha=1$ and 2 the shape is almost indistinguishable at
$\Delta=10$, see the bottom right panel. The trace length is $T=10^4$ and the
number of traces used for averaging is $10^3$.}
\label{fig_phi}
\end{figure}

Figure \ref{fig-tamsd-aged} shows the time averaged MSD $\overline{\delta^2(\Delta)
}$ of individual SBM traces for the case of weak, intermediate, and strong ageing
for different values of $\alpha$. We observe that the spread of individual
$\overline{\delta^2(\Delta)}$ changes only marginally with progressive ageing times
$t_a$. Also the changes with the scaling exponent $\alpha$ are modest, compare
figure \ref{fig_phi}.
Also note that the magnitude of the time averaged MSD decreases with $t_a$ for
ultraslow SBM at $\alpha=0$, stays independent on $t_a$ for Brownian motion at
$\alpha=1$, and increases with the ageing time for superdiffusive processes at
$\alpha>1$. These trends are in agreement with the theoretical predictions of
equation (\ref{eq-aged-sbm-delta-2-T-Delta}) shown as the solid lines in figure
\ref{fig-tamsd-aged}.

\section{Ergodicity breaking of non-ageing scaled Brownian motion}
\label{sec-model-simul}
\label{sec-non-aged}

\subsection{General expression for the ergodicity breaking parameter}

\begin{figure}
\includegraphics[width=16cm]{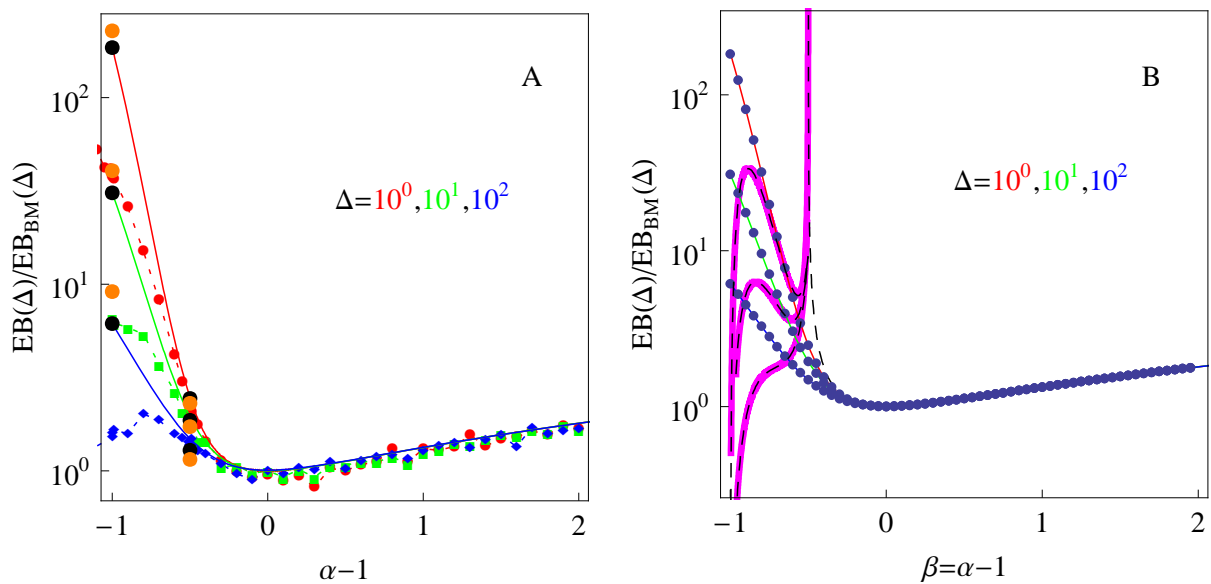}
\caption{Ergodicity breaking parameter EB of non-ageing SBM. (A) Results of
numerical simulations are depicted by the data points. The analytical results based
on equations (\ref{eq-eb-sbm-analyt}) and (\ref{eq-eb-sbm-analyt1}) are given by
the solid coloured lines. Data points for different lag times are shown in different
colours. The values  of EB for ultraslow SBM (\ref{eq-eb-usbm}) at $\alpha=0$ and at
$\alpha=1/2$ given by equation (\ref{eq-eb-1-over-2}) are shown as the bigger black
bullets, computed for $\Delta=10^0$, $10^1$, and $10^2$. The larger orange bullets
denote the same limits but without the additive constants to the leading functional
dependencies with $\Delta/T$. Parameters: the trace length is $T=10^4$, the number
of traces used for averaging at each $\alpha$ value is $N=10^3$. (B) Exact and
approximate analytical results for EB. The red, green, and blue curves are the exact
evaluations of equation (\ref{eq-eb-sbm-analyt}). The dashed curve in the region
$\alpha>1/2$ corresponds to equation (\ref{eq-eb-sbm}) and the dashed curves for
$0<\alpha<1/2$ are the results of \cite{soko14}. The magenta curves in the region
$0<\alpha<1/2$ are according to the analytical expansion (\ref{eq-eb-0-to-05}) for
given $\Delta$ values. The dark blue data points, coinciding with our exact
result (\ref{eq-eb-sbm-analyt}), follow from evaluating the double integral in
equation (9) of \cite{soko14} with Mathematica.}
\label{fig-eb}
\end{figure}

Analytically, the derivation of the EB parameter for SBM involves the evaluation
of the fourth order moment of the time averaged MSD,
\begin{eqnarray}
\nonumber
\left<\left(\overline{\delta^2(\Delta)}\right)^2\right>=
\frac{1}{T-\Delta}
\int\limits_{0}^{T-\Delta}dt_1\int\limits_0^{T-\Delta}dt_2&&
\left<\Big(x^2(t_1+\Delta)-x(t_1))^2\right.\\
&&\left.\times(x^2(t_2+\Delta)-x(t_2)\Big)^2\right>.
\label{eq-delta-4-general}
\end{eqnarray} 
We use the fundamental property of SBM that
\begin{equation}
\left<x(t_1)x(t_2)\right>=\left<x^2(\mathrm{min}\{t_1,t_2\})\right>,
\label{eq-pair-corr}
\end{equation}
and the Wick-Isserlis theorem for the fourth order correlators \cite{wick}.
We then obtain the nominator $\mathcal{N}$ of the EB parameter of equation
(\ref{eq-eb-via-xi})
\begin{eqnarray}
\nonumber
\mathcal{N}(\Delta)&=&\left<\left(\overline{\delta^2(\Delta)}\right)^2\right>-
\left<\overline{\delta^2(\Delta)}\right>^2\\
\nonumber
&=&\frac{2}{(T-\Delta)^2}\int_0^{T-\Delta}dt_1\int_0^{T-\Delta}dt_2 
\left<\Big(x(t_1+\Delta)-x(t_1))\right.\\
&&\hspace*{4cm}\times\left.(x(t_2+\Delta)-x(t_2)\Big)\right>^2.
\label{eq-eb-nominator-after-wick}
\end{eqnarray}
Taking the averages by help of equation (\ref{eq-pair-corr}) we arrive at 
\begin{eqnarray}
\nonumber
\mathcal{N}(\Delta)&=&\frac{4}{(T-\Delta)^2}\int_0^{T-\Delta}dt_1\int_{t_1}^{
T-\Delta}dt_2\left[\Big<x^2(t_1+\Delta)\Big>\right.\\
&&\hspace*{4cm}\left.-\Big<x(t_1+\Delta)x(t_2)\Big>\right]^2.
\label{eq-eb-nominator-after-wick-after-averagng}
\end{eqnarray}
With the new variable $\tau'=t_2-t_1$ (assuming $t_2>t_1$) and by changing the
order of integration we find the expression
\begin{eqnarray}
\nonumber
\mathcal{N}(\Delta)&=&\frac{4}{(T-\Delta)^2}\int_0^{\Delta}d\tau'\int_0^{T-
\Delta-\tau'}dt_1\left[\Big<x^2(t_1+\Delta)\Big>\right.\\
&&\hspace*{4cm}\left.-\Big<x^2(t_1+\tau')\Big>\right]^2.
\label{eq-eb-nominator}
\end{eqnarray}
Now, the new variables $x'=t_1/\Delta$ and $y'=\tau'/\Delta$ are introduced.
Substituting equation (\ref{msd}) into equation(\ref{eq-eb-nominator}) we obtain 
\begin{eqnarray}
\nonumber
\mathcal{N}(\Delta)&=&\frac{16 K_\alpha^2 \Delta^{2\alpha+2}}{(T-\Delta)^2}
\int_0^1dy'\int_0^{T/\Delta-1-y'}dx'\\
&&\times\left[(x'+1)^{2\alpha}-2(x'+1)^\alpha (x'+y')^\alpha+(x'+y')^{2\alpha}
\right].
\label{eq-eb-nominator-2}
\end{eqnarray}
Splitting the double integral over the variable $x'$ into an integral over a square
region and a triangular region yields
\begin{eqnarray}
\nonumber
&&\int_0^1dy'\int_0^{T/\Delta-2}dx'+\int_0^1dy'\int_{T/\Delta-2}^{T/\Delta
-1-y'}dx'\\
&&\hspace*{1.8cm}=\int_0^{T/\Delta-2}dx'\int_0^1dy'+\int_{T/\Delta-2}^{T/\Delta
-1}dx'\int_0^{T/\Delta-1-x'}dy'.
\end{eqnarray}
From the double integrals from the power-law functions in equation
(\ref{eq-eb-nominator-2}), via equation (\ref{eq-pair-corr}) we compute the
nominator as
\begin{eqnarray}
\label{eq-eb-sbm-analyt}
\nonumber
\mathcal{N}(\Delta,\tau)&=&\frac{16 K_\alpha^2 \Delta^{2\alpha+2}}{(T-\Delta)^2}
\left[\frac{(\tau-1)^{2\alpha+1}}{2\alpha+1}+\frac{(3\alpha+1)(\tau-1)^{2\alpha
+2}}{2(\alpha+1)^2(2\alpha+1)}\right.\\
\nonumber
&&-\frac{2\tau^{\alpha+1}(\tau-1)^{\alpha+1}}{(\alpha+1)^2}+\frac{\tau^{2\alpha
+2}}{2(\alpha+1)(2\alpha+1)}-\frac{(2\alpha^2+\alpha+1)}{2(\alpha+1)^2(2\alpha+1)}\\
&&\left.+\frac{2}{\alpha+1}\int_0^{\tau-1}dx'~(x')^{\alpha+1} (x'+1)^{\alpha}
\right],
\end{eqnarray}
in terms of the variable
\begin{equation}
\tau=\frac{T}{\Delta}.
\end{equation}
The integral
\begin{equation}
\label{integral}
I_1(\tau)=\int_0^{\tau-1}dx'~(x')^{\alpha+1}(x'+1)^{\alpha}
\end{equation}
remaining in the last term
of this expression can, in principle, be represented in terms of the incomplete
Beta-function.
The denominator $\mathcal{D}(\Delta)$ of the EB parameter (\ref{eq-eb-via-xi}) is
just the squared time averaged MSD given by equation (\ref{eq-sbm-tamsd}). We thus
arrive at the expression
\begin{eqnarray}
\label{eq-eb-sbm-analyt1}
\mathcal{D}(\Delta,\tau)&=&\left[\frac{2 K_\alpha \Delta^{\alpha+1}}{(\alpha+1)(
T-\Delta)}(\tau^{\alpha+1}-1-(\tau-1)^{\alpha+1})\right]^2.
\end{eqnarray}
Note that the double analytical integration of equation (9) in \cite{soko14} via
Wolfram Mathematica  yields a result, that is indistinguishable from equation
(\ref{eq-eb-sbm-analyt}), as demonstrated by the blue dots in figure \ref{fig-eb}B.

\subsection{Expansions and Limiting Cases}

We here consider some limiting cases of the EB parameter based on expressions
(\ref{eq-eb-sbm-analyt}) and (\ref{eq-eb-sbm-analyt}). In the limit $\alpha=1$
and for $\Delta/T\ll1$ the leading order expansion in terms of $\Delta/T$ turns
into equation (\ref{eq-eb-bm}). As it should the SBM process reduces to the
ergodic behaviour of standard Brownian motion.

\subsubsection{The case $0<\alpha<1/2$.}

The general expression for the behaviour of the EB parameter in the range $0<
\alpha<1/2$ follows from equation (\ref{integral}) by help of the identity
[equation (1.2.2.1) in Ref.~\cite{prud}]
\begin{equation}
\int x^p(x+1)^qdx=\frac{x^{p+1}(x+1)^q}{p+q+1}+\frac{q}{p+q+1}\int x^p(x+1)^{
q-1}dx,
\label{eq-prud-formula}
\end{equation} that can be checked by straight differentiation. Performing this
sort of partial integration three times we reduce the power of the integrand so
that in the limit $\tau\to\infty$ the integral becomes a converging function.
In the range $0<\alpha<1/2$ we the find exact expression
\begin{eqnarray}
\nonumber
I_1(\tau)&=&\frac{(\tau-1)^{\alpha+2}\tau^\alpha}{2(\alpha+1)}+
\frac{\alpha(\tau-1)^{\alpha+2}\tau^{\alpha-1}}{2(\alpha+1)(2\alpha+1)}
+\frac{\alpha(\alpha-1)(\tau-1)^{\alpha+2}\tau^{\alpha-2}}{4\alpha(\alpha+1)
(2\alpha+1)}\\
&&+\frac{\alpha(\alpha-1)(\alpha-2)}{4\alpha(\alpha+1)(2\alpha+1)}
\times\int_0^{\tau-1} (x')^{\alpha+1}(x'+1)^{\alpha-3}dx'.
\label{I1-alfa-smaller-05}
\end{eqnarray}
The remaining converging integral can be represented in the limit $\Delta/T\ll1$
via the Beta function: setting the upper integration limit $(\tau-1) \to \infty$
we obtain 
\begin{equation}
\int_0^{\infty}(x')^{\alpha+1}(x'+1)^{\alpha-3}dx'=B(\alpha+2,1-2\alpha).
\end{equation}
Then we arrive at the following scaling law for the EB parameter,
\begin{equation}
\mathrm{EB}(\alpha,\Delta)\sim4C(\alpha)\left(\frac{\Delta}{T}\right)^{2\alpha},
\label{eq-eb-0-to-05}
\end{equation}
where the coefficient is given by
\begin{equation}
C(\alpha)=\frac{(1-\alpha)(2-\alpha)B(\alpha+2,1-2\alpha)-(2\alpha^2+\alpha+1)}
{2(\alpha+1)^2(2\alpha+1)}.
\label{eq-calfa-coeff}
\end{equation} 
The scaling form of EB versus $(\Delta/T)$ of equation (\ref{eq-eb-0-to-05})
coincides with that proposed in reference \cite{soko14}, and it is indeed valid
for vanishing $\Delta/T$ and scaling exponents not too close to $\alpha=0$ and
$\alpha=1/2$, see below. We find in addition that in the region $0<\alpha\lesssim
1/2$ the EB parameter of the SBM process becomes a sensitive function of the lag
time $\Delta$, as shown in figure \ref{fig-eb}A, both from our theoretical results
and computer simulations. This means that no universal rescaled variable $\Delta/T$
exists, as is the case for standard Brownian motion. 

The asymptote (\ref{eq-eb-0-to-05}) agrees with the result (10) in \cite{soko14} 
in the range $0<\alpha<1/2$ of the scaling exponent and for infinitely large values
$\tau$. Equation (\ref{eq-calfa-coeff}) above provides an explicit form for the 
prefactor. In figure \ref{fig-eb}B the approximate expansion (\ref{eq-eb-0-to-05})
is shown as magenta curve. At realistic values $\Delta/T$ the asymptote
(\ref{eq-eb-0-to-05}) agrees neither with our exact expression
(\ref{eq-eb-sbm-analyt}) nor with the simulation data. As this demonstrates the
exact expression (\ref{eq-eb-sbm-analyt}) needs to be used a forteriori. The main
reason is the finite $\tau$ value used in the simulations: for very small $\Delta/T$
equation (\ref{eq-eb-0-to-05}) describes the exact result (\ref{eq-eb-sbm-analyt})
significantly better (not shown). We note that away from the critical points at
$\alpha=0$ and $\alpha=1/2$, equation (\ref{eq-eb-0-to-05}) returns zero and
infinity, respectively (magenta curves in figure \ref{fig-eb}B). At these points
special care is required when computing $I_1$ in equation (\ref{I1-alfa-smaller-05}),
as discussed below.

\subsubsection{The case $\alpha>1/2$.}

For values $\alpha>1/2$ of the scaling exponent in the limit of small $\Delta/T$ 
the denominator (\ref{eq-eb-sbm-analyt1}) becomes $\mathcal{D}(\tau)\simeq4\tau^{2
\alpha}$. Note that here we need to include two more iterations of the integral
in the last term of equation (\ref{I1-alfa-smaller-05}) by using equation
(\ref{eq-prud-formula}). Then we arrive at a new integral term that is converging
at $\tau\to\infty$. Thus the nominator (\ref{eq-eb-sbm-analyt})---after cancellation
of the first three orders in the expansion in terms of large $\tau$---yields to
leading order $\mathcal{N}(\tau)\simeq16\alpha^2\tau^{2\alpha-1}/[3(2\alpha-1)]$.

From the exact expression (\ref{eq-eb-sbm-analyt}) by using the integration
formula (\ref{eq-prud-formula}) four times, we find the exact representation
\begin{eqnarray}
\nonumber
I_1(\tau)&=&\frac{(\tau-1)^{\alpha+2}\tau^\alpha}{2(\alpha+1)}+
\frac{\alpha(\tau-1)^{\alpha+2}\tau^{\alpha-1}}{2(\alpha+1)(2\alpha+1)}+
\frac{\alpha(\alpha-1)(\tau-1)^{\alpha+2}\tau^{\alpha-2}}{4\alpha(\alpha+1)
(2\alpha+1)}\\ 
\nonumber
&&+\frac{\alpha(\alpha-1)(\alpha-2)(\tau-1)^{\alpha+2}\tau^{\alpha-3}}{4\alpha
(\alpha+1)(2\alpha+1)(2\alpha-1)}\\
&&+\frac{\alpha(\alpha-1)(\alpha-2)(\alpha-3)}{4\alpha(\alpha+1)(2\alpha+1)
(2\alpha-1)}\times\int_0^{\tau-1}(x')^{\alpha+1}(x'+1)^{\alpha-4}dx'.
\label{I1-alfa-greater-05-2}
\end{eqnarray}
From this expression the leading term with the divergence at $\alpha=1/2$ is
written explicitly and the remaining integral is converging only then. Plugging
this expression into equations (\ref{eq-eb-sbm-analyt}) and (\ref{eq-eb-sbm-analyt1})
and keeping terms of order $\tau^{2\alpha-1}$ in the limit $\tau\gg1$ we recover
the result of \cite{soko14} given by equation (\ref{eq-eb-sbm}), again valid in
the range $\alpha>1/2$. Note that the divergence in the denominator of the last
term in $I_1$ in equation (\ref{I1-alfa-greater-05-2}) is compensated by the
proper expansion of the remaining integral in $I_1$ in the limit of large values
of $\tau$ for $\alpha>1/2$, see below.
 
The EB parameter then scales as 
\begin{equation}
\lim_{\Delta/T\to0}\mathrm{EB}(\Delta)\sim\frac{4}{3}\frac{\alpha^2}{2\alpha-1}
\frac{\Delta}{T}.
\label{eq-eb-sbm}
\end{equation}
This result coincides with expression (10) in \cite{soko14} in the range $\alpha
>1/2$. As mentioned already, special care is needed near the critical point
$\alpha=1/2$. Equation (\ref{eq-eb-sbm}) implies that SBM is an ergodic process, 
with the EB parameter scaling strictly linearly with $\Delta/T$ as in relation
(\ref{eq-eb-bm}) for Brownian motion, however, with an $\alpha-$dependent prefactor
of the form $\alpha^2/(2\alpha-1)$. In contrast to subdiffusive CTRW processes
\cite{metz14,metz08} and heterogeneous diffusion processes \cite{cher13} the EB
parameter for Brownian motion converges to zero and thus for sufficiently long
measurement times the result of time averaged observables become reproducible.

\subsubsection{The case $\alpha=0$}

Now let us focus on the critical points $\alpha=0$ and $\alpha=1/2$ in detail. At
$\alpha\to0$ the EB parameter of the ultraslow SBM process \cite{bodr15a} can be  
obtained from equation (\ref{eq-eb-sbm-analyt}). To this end we first expand
result (\ref{eq-eb-sbm-analyt}) for small $\alpha$ using the identity $x^\alpha=
e^{\alpha\log(x)}$. In the remaining integral $I_1$ in equation
(\ref{integral}) we first expand the integrand in powers of small $\alpha$
and then integrate the expanded function in the limits $\int_0^{T-\Delta}dt$. The
first two orders of the expansion in $\alpha$ in the nominator of EB disappear.
Dividing the leading orders in $\alpha^2$ in the nominator and denominator of EB
and expanding for short lag times $\Delta/T\ll1$ afterwards to the leading order 
we find
\begin{equation}
\lim_{\Delta/T \to 0}\mathrm{EB}_\mathrm{USBM}(\Delta)\sim\frac{4(\pi^2/6-1)}{
(\log[T/\Delta]+1)^2}.
\label{eq-eb-usbm}
\end{equation} 
This result was obtained from independent considerations for ultraslow SBM as
equation (20) in \cite{bodr15a}. Note the logarithmic rather than the linear
dependence of EB on $\Delta/T$ in this case, stemming from the ultraslow
logarithmic scaling of the MSD and the time averaged MSD with (lag) time.

\subsubsection{The case $\alpha=1/2$}

Similarly, to explore the limit $\alpha\to1/2$ we first expand the exact result
(\ref{eq-eb-sbm-analyt}) for $\mathcal{N}(\Delta)$ in $\alpha$ around this point.
In analogy to the case $\alpha=0$ we expand the integrand in $I_1$ in terms of
powers of $(\alpha-1/2)$ and then perform the integration over $t$ from $0$ to
$T-\Delta$. Dividing the expansion of the nominator (\ref{eq-eb-sbm-analyt}) of
EB, taken at $\alpha=1/2$\footnote{With regard to the higher order expansion
taken below, this corresponds formally to an expansion of order $(\alpha-1/2)^0$.}
in the limit $\Delta/T\to0$, by the leading
order of the denominator (\ref{eq-eb-sbm-analyt1}) in the same limit---scaling as
$4\tau$---we get 
\begin{equation}
\lim_{\Delta/T\to0}\mathrm{EB}_{\alpha=1/2}(\Delta)=\frac{\Delta}{3T}\Big[\log(T
/\Delta)+2 \log(2)-5/6\Big].
\label{eq-eb-1-over-2}
\end{equation}
The same expression can be obtained by expanding equation (\ref{I1-alfa-smaller-05}) 
valid in the region $0<\alpha<1/2$. Alternatively result (\ref{eq-eb-1-over-2}) can
be obtained from the exact expression (\ref{I1-alfa-greater-05-2}) valid for
$\alpha>1/2$. In this case, however, due to a pole at $\alpha=1/2$ one more order
in the power expansion near $\alpha=1/2$ needs to be properly evaluated when
expanding $I_1$. Then, the divergence in the denominator of the prefactor of the 
last term in equation (\ref{I1-alfa-greater-05-2}) becomes eliminated and the EB
parameter stays continuous as $\alpha\to1/2$.

Compared to the case $\alpha=1$ of Brownian motion the result (\ref{eq-eb-1-over-2})
for EB features a weak logarithmic dependence on $\Delta/T$. As expected
the values of EB according to equation (\ref{eq-eb-1-over-2}) are very close to the 
exact solution (\ref{eq-eb-sbm-analyt}), as shown by the larger black bullets for
$\alpha
=1/2$ in figure \ref{fig-eb}A. Note that for finite $T/\Delta$ values the additional
constants following the leading functional dependencies in equation
(\ref{eq-eb-usbm}) and equation (\ref{eq-eb-1-over-2}) play a significant r{\^o}le,
as seen in figure \ref{fig-eb}A. The agreement of these EB values with the exact 
predictions of equation (\ref{eq-eb-sbm-analyt}) and computer simulations is
particularly good for smaller $\Delta/T$ values, as expected based on the large
$\tau$ expansions used in the derivation of equations (\ref{eq-eb-usbm}) and
(\ref{eq-eb-1-over-2}).

\subsection{Computer Simulations}

We implement the same algorithms for the iterative computation of the particle
displacement $x(t)$ as developed for the heterogeneous diffusion process
\cite{cher13} and the combined heterogeneous diffusion-scaled Brownian motion
process \cite{cher15b}. We simulate the one dimensional overdamped Langevin
equation 
\begin{equation}
\frac{dx(t)}{dt}=\sqrt{2D(t)}\times\xi(t)
\label{eq-dc}
\end{equation}
driven by the Gaussian white  noise $\xi(t)$ of unit intensity and zero mean.
At step $i+1$ the particle displacement is  
\begin{equation}
\label{eq-simul-scheme}
x_{i+1}-x_i=\sqrt{2[D(t_i)+C]}(y_{i+1}-y_{i}),
\end{equation}
where the increments $(y_{i+1}-y_i)$ of the Wiener process represent a $\delta$
correlated Gaussian noise with unit variance and zero mean. Unit time intervals
separate consecutive iteration steps. To avoid a possible particle trapping at
the pole of $D(t)$ we introduced the small constant $C=10^{-3}$ in analogy to the
procedure for heterogeneous diffusion processes \cite{cher13}. The initial position
of the particle is $x_0=x(t=0)=0.1$.

Our simulations results shown in figure \ref{fig-eb}A confirm the validity of the
general analytical expressions (\ref{eq-eb-sbm-analyt}) and (\ref{eq-eb-sbm-analyt1})
making up the EB parameter in the whole range of the scaling exponent $\alpha$.
We also find that the short lag time expansion (\ref{eq-eb-sbm}) agrees well with
the exact solution and simulations at $\alpha\gtrsim1/2$ (figure \ref{fig-eb}B).
In the range $\alpha\gtrsim1/2$ the EB parameter for $\Delta/T\ll1$ is nearly
insensitive to the lag time and grows with $\alpha$ in accord with equation
(\ref{eq-eb-sbm}). In particular, the full analytical expression for EB (equations
(\ref{eq-eb-sbm-analyt}) and (\ref{eq-eb-sbm-analyt1})) and the
results of the simulations show no divergence at $\alpha=1/2$, in contrast to the
approximate results of reference \cite{soko14}.

Figure \ref{fig-eb}A also shows the approximate EB values (\ref{eq-eb-usbm}) for
ultraslow SBM as well as EB at $\alpha=1/2$ from equation (\ref{eq-eb-1-over-2})
indicated as larger points. These points are close to our predictions for SBM 
at $\alpha\to0$, in particular, for small $\Delta/T$ values when the approximations
used in deriving the corresponding equations are better satisfied. As the ratio
$\Delta/T$ grows and the scaling exponent converges to zero, $\alpha\to
0$---indicating progressively slower diffusion---the results of our simulations
start to deviate from the exact analytical results (\ref{eq-eb-sbm-analyt}) and
(\ref{eq-eb-sbm-analyt1}), as shown in figure \ref{fig-eb}. In this limit
apparently better statistics are needed in the simulations.

\begin{figure}
\begin{center}
\includegraphics[width=12cm]{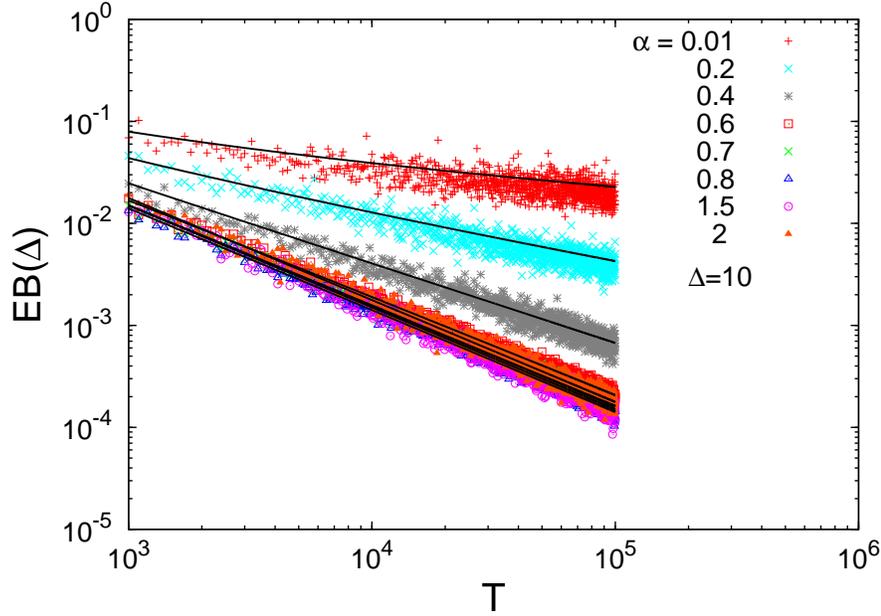}
\end{center}
\caption{EB parameter for non-ageing SBM versus trace length $T$. The solid lines
represent the exact results according to equation (\ref{eq-eb-sbm-analyt}).
Parameters: $\Delta=10$ and $N=10^3$.}
\label{fig-eb-t}\end{figure}

In figure \ref{fig-eb-t} we show that EB scales with the trace length $T$
approximately as $1/T^{2\alpha}$ for $0<\alpha<1/2$ and as $1/T$ for $\alpha>1/2$;
compare to the results in figure 1 of reference \cite{soko14}.

\section{Ergodicity breaking of ageing scaled Brownian motion}
\label{sec-aged}

We consider the ergodic properties of ageing SBM, where $t_a$ denotes the time
span in between the initiation of the system and start of the measurement.
The ergodicity breaking parameter is defined through the ageing time averaged
MSD (compare equations (\ref{eq-TAMSD-aged}) and (\ref{eq-aged-sbm-delta-2-T-Delta}))
as
\begin{eqnarray}
\label{eq-EB-aged-general}
\mathrm{EB}_a(\Delta)=\frac{\left\langle\overline{\delta^{2}_a(\Delta)}^2\right
\rangle-\left\langle \overline{\delta^2_a(\Delta)}\right\rangle^2}{\left\langle
\overline{\delta^2_a(\Delta)}\right\rangle^2}=\frac{\mathcal{N}_a(\Delta,\tau)}{
\mathcal{D}_a(\Delta,\tau)}
\label{eq-eq-EB-sbm-final-aged}
\end{eqnarray}
For the numerator we find in full analogy to the non-ageing situation
\begin{eqnarray}
\nonumber
\mathcal{N}_a(\Delta)&=&\frac{4}{(T-\Delta)^2}  
\int_{t_a}^{T+t_a-\Delta} dt_1 \int_{t_1}^{T+t_a-\Delta} dt_2\\
&&\times\left[ \left\langle x^2 (t_{1}+\Delta)\right\rangle - 
\left\langle x(t_1+\Delta)x(t_2)\right\rangle\right]^2.
\label{eq-EB-via-t1-t2}
\end{eqnarray}
Changing the variables as above for the non-ageing scenario, $\tau'=t_2-t_1$, 
we switch the limits of integration using $t_1(\tau')=T+t_a-\Delta-\tau'$ and then
split the integrals over $\tau'$ to compute the pair correlators using the property
(\ref{eq-pair-corr}). This yields the representation of the nominator of EB in terms
of one-point averages only,
\begin{eqnarray}
\nonumber
\mathcal{N}_a(\Delta)&=&\frac{4}{(T-\Delta)^2}\\
&&\times\int_0^{\Delta}d\tau'\int_{t_a}^{T+t_a-\Delta-\tau'}dt_1  
\Big[\left\langle x^2 (t_{1}+\Delta)\right\rangle - 
\left\langle x^2(t_{1}+\tau')\right\rangle\Big]^2.
\label{eq-nominator-via-MSDs-only}
\end{eqnarray}
We proceed by inserting the MSDs of equation (\ref{msd}) and arrive at 
\begin{eqnarray}
\nonumber
\mathcal{N}_a(\Delta)&=&\frac{16K_{\alpha}^2 \Delta^{2\alpha+2}}{(T-\Delta)^{2}}
\int_0^1dy'\int_{{t_a}/{\Delta}}^{T/{\Delta}+t_{a}/{\Delta}-1-y'}dx'\\
&&\times\Big[(x'+1)^{2\alpha}-2(x'+1)^{\alpha}(x'+y')^{\alpha} +(x'+y')^{2\alpha}
\Big].
\label{eq-nomi-EB-aged-via-x-y-power-alpha}
\end{eqnarray}
Changing the order of integration and splitting the integral over $x'$ we get in
terms of the variables $\tau=T/\Delta$ and
\begin{equation}
\tau_a=\frac{t_a}{\Delta}
\end{equation}
that
\begin{eqnarray}
\label{eq-EB-aged-2-integrals}
\nonumber
\mathcal{N}_a(\Delta,\tau)&=&\frac{16K_{\alpha}^2 \Delta^{2\alpha+2}}{(T-\Delta)^2}
\int_{\tau_a}^{\tau+\tau_a-2} dx'\int_0^1dy'\Big[(x'+1)^{2\alpha}-2(x'+1)^{\alpha}
(x'+y')^{\alpha}\\
\nonumber
&&+(x'+y')^{2\alpha}\Big]
+\int_{\tau+\tau_a-2}^{\tau+\tau_a-1}dx'\int_0^{\tau+\tau_a-1-x'}dy'\\  
&&\times\Big[(x'+1)^{2\alpha}-2(x'+1)^{\alpha}(x'+y')^{\alpha}+(x'+y')^{
2\alpha}\Big].
\end{eqnarray}
Finally, taking the integrals in the nominator of EB for ageing SBM yields
\begin{eqnarray}
\nonumber
\mathcal{N}_a(\Delta,\tau)&=&\frac{16K_{\alpha}^{2}\Delta ^{2\alpha+2}}{
(T-\Delta)^{2}}\left[\frac{(\tau+\tau_{a}-1)^{2\alpha+1}}{2\alpha+1}
-\frac{(\tau_a+1)^{2\alpha+1}}{2\alpha+1}\right.\\
\nonumber
&&+\frac{(3\alpha+1)(\tau+\tau_a-1)^{2\alpha+2}}{2(2\alpha+1)(\alpha+1)^2}
+\frac{(3\alpha+1)(\tau_a+1)^{2\alpha+2}}{2(2\alpha+1)(\alpha+1)^2}\\
\nonumber
&&+\frac{(\tau_a)^{2\alpha+2}}{2(2\alpha+1)(\alpha+1)}\\
\nonumber
&&+\frac{(\tau+\tau_a)^{2\alpha+2}}{2(2\alpha+1)(\alpha+1)}
-\frac{2(\tau+\tau_a)^{\alpha+1}(\tau+\tau_a-1)^{\alpha+1}}{(\alpha+1)^2} \\
&&\left.+\frac{2}{\alpha+1}\int_{\tau_a}^{\tau+\tau_a-1}dx'~(x')^{\alpha+1} (x'+1)^
\alpha\right].
\label{eq-eq-sbm-final-nominator-aged}
\end{eqnarray}
Here we again denote
\begin{equation}
I_1(\tau,\tau_a)=\int_{\tau_a}^{\tau+\tau_a-1}dx' (x')^{\alpha+1}(x'+1)^\alpha.
\end{equation}
The denominator of EB follows from the time averaged MSD
(\ref{eq-aged-sbm-delta-2-T-Delta}), namely \cite{cher15b,metz15}
\begin{eqnarray}
\nonumber
\mathcal{D}_a(\Delta,\tau)&=&\left<\overline{\delta^2_a(\Delta)}\right>^2=
\Big(\frac{2K_{\alpha}\Delta^{\alpha+1}}{(\alpha+1)(T-\Delta)}
\Big[(\tau+\tau_a)^{\alpha+1}-(\tau_a+1)^{\alpha+1}\\
&&-(\tau+\tau_a-1)^{\alpha+1}+\tau_a^{\alpha+1}\Big]\Big)^2.
\label{eq-aged-sbm-denomi-via-tau}
\end{eqnarray}
The final EB breaking parameter (\ref{eq-eq-EB-sbm-final-aged}) for ageing SBM
turns into expression (\ref{eq-eb-sbm-analyt}) for the non-ageing case, $\tau_a=0$.

\begin{figure}
\begin{center}
\includegraphics[width=12cm]{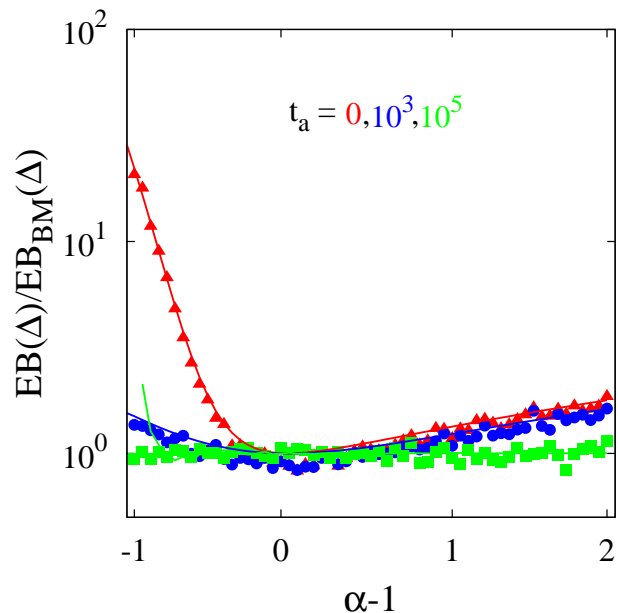}
\end{center}
\caption{EB parameter for ageing SBM. Results of simulations are shown by the
points and the analytical results (\ref{eq-eq-sbm-final-nominator-aged}) are
represented by the solid lines of the corresponding colour. Parameters: $\Delta
=10$, $T=10^4$, and $N=10^3$.}
\label{fig-eb-aged}
\end{figure}

In the limit of strong ageing, $\tau_a\gg T\gg\Delta$, the time averaged MSD scales
as 
\begin{equation}
\left<\overline{\delta^2_a(\Delta)}\right>\sim2\alpha K_\alpha t_a^{\alpha-1}
\Delta
\label{eq-tamsd-strong-age}
\end{equation}
and the nominator of EB grows as
\begin{equation}
\mathcal{N}_a(\Delta,\tau)\sim16 K_{\alpha}^{2}\Delta ^{2\alpha}\tau^{-2} 
(\alpha^2\tau_a^{2\alpha-2}\tau/3)
\end{equation}
to leading order in large $\tau_a$ values and long trajectories. Then, the ergodicity 
breaking parameter follows the Brownian law (\ref{eq-eb-bm}). This limiting
behaviour is supported by the simulations of strongly ageing SBM shown in figure
\ref{fig-eb-aged}. Moreover, it is similar to that of ageing ultraslow SBM
\cite{bodr15a}. Physically, in the limit of long ageing times $\tau_a$ the 
diffusivity $D(t)$ changes only marginally on the time scale $T\ll t$ of the
particle diffusion, so that the entire process stays approximately ergodic.

In the opposite limit of weak ageing, $\tau_a\ll T$, we observe that $\left<\overline{
\delta^2_a(\Delta,\tau)}\right>\sim2K_\alpha\Delta^\alpha(\tau^{\alpha-1}+\alpha
\tau_a\tau^{\alpha-2})$, and the nominator of EB to leading order of short $\tau_a$
and long $T$ values produces $\mathcal{N}(\Delta,\tau)\sim 16 K_{\alpha}^2\Delta
^{2\alpha}\tau^{-2}(\alpha^2\tau^{2\alpha-1}/[3(2\alpha-1)])$. Consequently the
EB parameter to leading order is independent of the ageing time $\tau_a$ and follows
equation (\ref{eq-eb-sbm}) as long as $\alpha>1/2$. 

\begin{figure}
\begin{center}
\includegraphics[width=12cm]{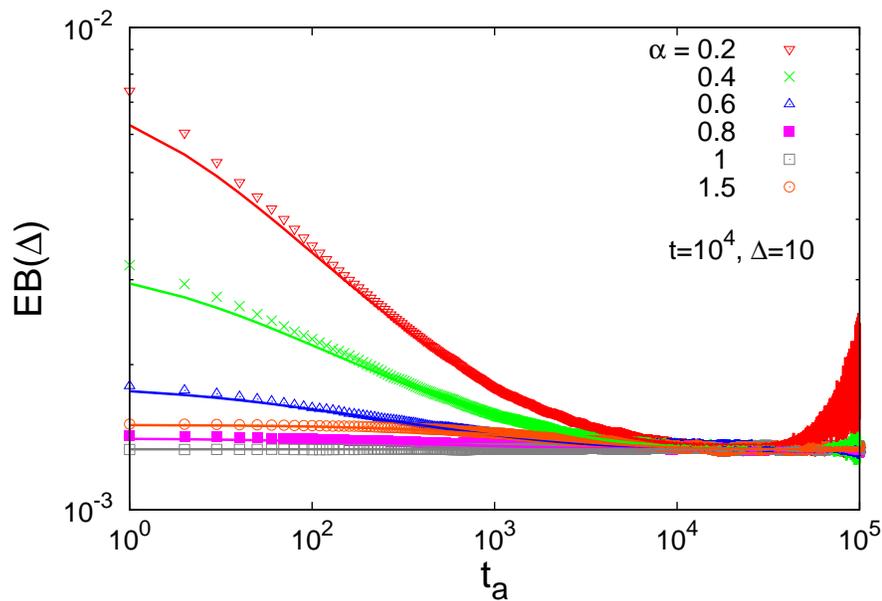}
\end{center}
\caption{EB parameter for ageing SBM versus ageing time $t_a$. Analytical results
(\ref{eq-eq-sbm-final-nominator-aged}) and (\ref{eq-aged-sbm-denomi-via-tau}) for
different $\alpha$ values are represented
by the solid lines. Some instabilities in the simulations are visible at long ageing
times, in particular for small $\alpha$. Parameters: $\Delta=10$, $T=10^4$, and
$N=10^3$.}
\label{fig-eb-ta}
\end{figure}

Figure \ref{fig-eb-aged} shows the simulations results based on the stochastic
Langevin process of ageing SBM. We find that in the limit of strong ageing, 
consistent with our theoretical results the EB of ageing SBM indeed approaches
the Brownian limit (\ref{eq-eb-bm}). For weak and intermediate ageing the general
EB expression (\ref{eq-eq-sbm-final-nominator-aged}) is in good agreement with the
simulations results, compare the data sets in figure \ref{fig-eb-aged}. Finally
figure \ref{fig-eb-ta} depicts the graph of EB versus ageing time explicitly, 
together with the theoretical results
(\ref{eq-eq-sbm-final-nominator-aged}) and (\ref{eq-aged-sbm-denomi-via-tau}).
We observe
that EB decreases with the ageing time and this reduction is particularly
pronounced for strongly subdiffusive SBM processes. The latter also feature some
instabilities upon the numerical solution of the stochastic equation for long
ageing times.

\section{Conclusions}
\label{sec-disc}

We here studied in detail the ergodic properties of SBM with its power-law time
dependent diffusivity $D(t)\simeq t^{\alpha-1}$. In particular, we derived the
higher order time averaged moments and obtained the ergodicity breaking parameter
of SBM, which quantifies the degree of irreproducibility of time averaged
observables of a stochastic process. For the highly non-stationary,
out-of-equilibrium SBM process we analysed the EB parameter with respect to the 
scaling exponent $\alpha$, the lag time $\Delta$, and the trace length $T$. We
revealed a non-monotonic dependence $\mathrm{EB}(\alpha)$. In particular, we showed
that there is no divergence at $\alpha=1/2$, in contrast to the approximate results
of \cite{soko14}. We also obtained a peculiar dependence for the EB dependence
on the trace length $T$, $\mathrm{EB}(T)\sim1/T^{2\alpha}$ for $0<\alpha<1/2$ and
$\mathrm{EB}(T)\sim1/T$ for $\alpha>1/2$, in agreement with \cite{soko14}. 
We also obtained analytical and numerical results for EB for ageing SBM as function
of the model parameters and the ageing time $t_a$.

Our exact analytical results are fully supported by stochastic simulations. We find
that over the range $\alpha\gtrsim1/2$ and for $\Delta/T\ll1$ the EB dependence on
the lag time and trace length involves the universal variable $1/\tau=\Delta
/T$, as witnessed by equation (\ref{eq-eb-sbm}). For arbitrary lag times and trace
lengths the general result for ageing and non-ageing SBM are, however, more complex,
see equations (\ref{eq-eb-sbm-analyt}) and (\ref{eq-eq-sbm-final-nominator-aged}).
These are the main results of the current work.
For strongly subdiffusive SBM in the range of exponents
$0<\alpha\lesssim1/2$ the ergodic properties are, in contrast, strongly dependent
on the lag time $\Delta$. The correct limit of our exact result
(\ref{eq-eb-sbm-analyt}) was obtained for the EB parameter of ultraslow SBM with
$\alpha\to0$ and for SBM with exponent $\alpha=1/2$. Although EB has some additional
logarithmic scaling at this point, it reveals no divergence as $\alpha=1/2$ is
approached.

We are confident that the strategies for obtaining higher order time averaged
moments developed herein will be useful for the analysis of other anomalous
diffusion processes, in particular for the analysis of finite time corrections
of EB for fractional Brownian motion \cite{deng09} or for processes with
spatially and temporally random diffusivities \cite{lapeyre,gary}.

\ack

We acknowledge funding from the Academy of Finland (Suomen Akatemia, Finland
Distinguished Professorship to RM), the 
Deutsche Forschungsgemeinschaft (to AGC, IMS and FT), and the IMU Berlin Einstein
Foundation (to AVC).

\end{document}